\documentclass[twocolumn]{aastex61}
\usepackage{graphicx}
\usepackage{multirow}
\usepackage{textcomp}
\usepackage{epsfig}
\usepackage{amsmath}
\usepackage{amssymb}
\usepackage{amsthm}
\usepackage{natbib}

\def\ergs {erg\,s$^{-1}$}
\def\ergscm2 {erg\,s$^{-1}$cm$^{-2}$}

\def\cm2 {cm$^{-2}$}

\def\deg{$^{\circ}$}

\submitjournal{ApJ}

\shorttitle{Dust RT modelling of the SGR1900+14 infrared ring}
\shortauthors{G. Natale et al.}

\begin{document}

\title{Dust radiative transfer modelling of the infrared ring around the magnetar SGR\,1900$+$14}

\correspondingauthor{G. Natale}
\email{gnatale@uclan.ac.uk}

\author{G. Natale}
\affil{Jeremiah Horrocks Institute, University of Central Lancashire, Preston, PR1 2HE, UK}

\author{N. Rea}
\affiliation{Institute of Space Sciences (IEEC--CSIC), Campus UAB, Carrer de Can Magrans S/N, 08193 Barcelona, Spain.}
\affiliation{Anton Pannekoek Institute for Astronomy, University of Amsterdam, Postbus 94249, NL--1090 GE Amsterdam, the Netherlands.}

\author{D. Lazzati}
\affiliation{Department of Physics, Oregon State University, 301 Weniger Hall, Corvallis, OR 97331, USA}

\author{R. Perna}
\affiliation{Department of Physics and Astronomy, Stony Brook University, Stony Brook, NY, 11794, USA}

\author{D. F. Torres}
\affiliation{Institute of Space Sciences (IEEC--CSIC), Campus UAB, Carrer de Can Magrans S/N, 08193 Barcelona, Spain.}
\affiliation{Instituci\'o Catalana de Recerca i Estudis Avan\c{c}ats (ICREA), E-08010 Barcelona, Spain}

\author{J. M. Girart}
\affiliation{Institute of Space Sciences (IEEC--CSIC), Campus UAB, Carrer de Can Magrans S/N, 08193 Barcelona, Spain.}

\begin{abstract}
A peculiar infrared ring-like structure was discovered by  {\em Spitzer}  around the strongly magnetised neutron star  SGR\,1900$+$14. This infrared structure was suggested to be due to a dust-free cavity, produced by the SGR Giant Flare occurred in 1998, and kept illuminated by surrounding stars. Using a 3D dust radiative transfer code, we aimed at reproducing the emission morphology and the integrated emission flux of this structure assuming different spatial distributions and densities for the dust, and different positions for the illuminating stars.  We found that a dust-free ellipsoidal cavity can reproduce the shape, flux, and spectrum of the ring-like infrared emission, provided that the illuminating stars are inside the cavity and that the interstellar medium has high gas density ($n_H\sim$1000\,cm$^{-3}$). We further constrain the emitting region to have a sharp inner boundary and to be significantly extended in the radial direction, possibly even just a cavity in a smooth molecular cloud. We discuss possible scenarios for the formation of the dustless cavity and the particular geometry that allows it to be IR-bright.

\end{abstract}

\keywords{X-rays: stars --- stars: neutron --- stars:
individual (SGR\,1900$+$14) -- dust -- radiative transfer}

\section{Introduction}

Strongly magnetized neutron stars \citep[magnetars\\][]{Duncan92, Thompson93} are extremely powerful X-ray and soft gamma-ray emitters, in particular under the form of large flares. These flares might reach luminosities that in our Galaxy second only Supernova explosions. In particular, magnetars emit a large variety of flares and outbursts on timescales from fraction of seconds to years, on a vast range of luminosities from $10^{38}$ to $\sim10^{47}$\ergs \citep{Rea11, Turolla15}. The most energetic events they ever emitted, called {\em Giant Flares}, have been detected three times in the past few decades, from three magnetars: the Soft Gamma-ray Repeaters (SGR) 0526-66 on 1979 March 5 \citep{Mazets79}, SGR\,1900+14 on 1998 August 27 \citep{Hurley99} and the last and more energetic one on 2004 December 27 from SGR\,1806-20 \citep{Hurley05}. All of them are characterized by a very luminous initial spike ($\sim10^{44-47}$\ergs ), lasting less than a second, which decays rapidly into a softer tail (modulated at the neutron star spin period) lasting several hundreds of seconds (with luminosities of $\sim10^{43}$\ergs ). The nature of the steady and flaring high energy emission from these sources has been intriguing all along. In fact, magnetar's X-ray luminosity is in general too high to be produced by the pulsar rotational energy losses alone, as for more common isolated radio pulsars, and the lack of any companion star excludes an accretion scenario. It is now well established that the peculiarities of these extreme highly magnetized objects ($10^{14-15}$\,Gauss) are related to the strength and instability of their magnetic field, that at times might stress the stiff neutron star crust \citep{Thompson95, Perna11}, might rearrange itself locally in small twisted bundles, or disrupt and reconnect higher up in the magnetosphere producing large ejections of particles. The ages of the $\sim$25 magnetars known \citep[see][]{Olausen14}, derived from their rotational properties ($t_c\sim\dot{P}/P$), indicate a young population, typically a few thousand years old. In three or four cases there are reasonably accepted associations with Supernova Remnants \citep{Gaensler01}, as well as massive star clusters \citep{Muno06, Eikenberry01, Vrba00}.

\begin{figure*}[t]
\includegraphics[scale=0.7]{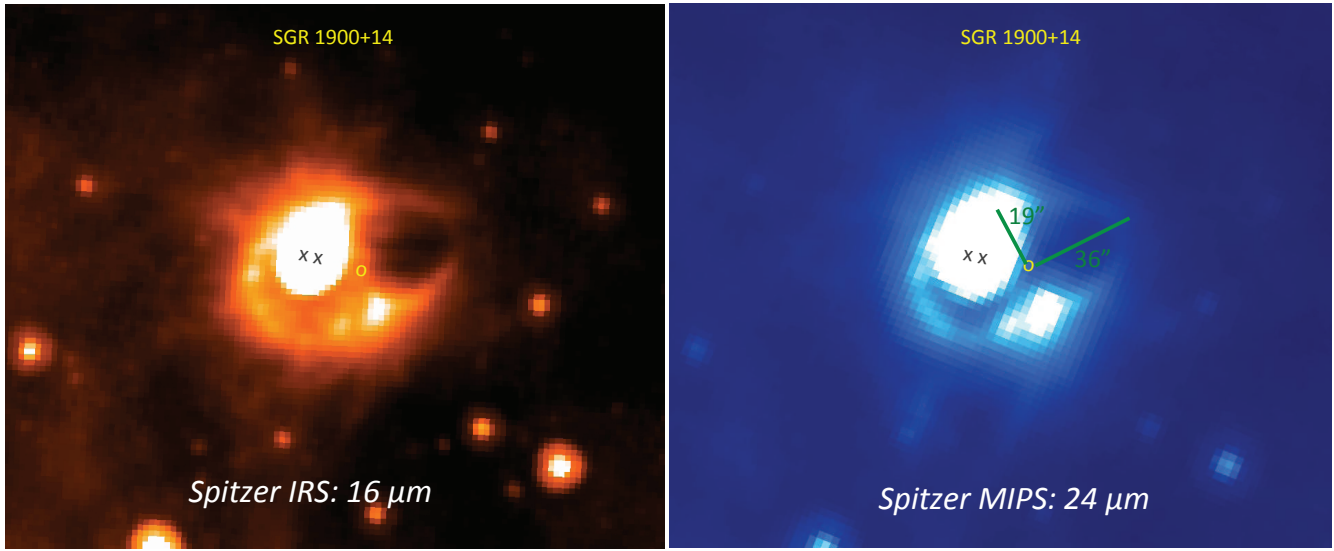}
\caption{{\em Spitzer} observations of SGR\,1900$+$14 at 16 and 24$\mu$m. The yellow circle shows the position of the SGR, while the two black crosses mark the positions of the two M stars \citep[for details on these observations see][]{Wachter08}.}
\label{spitzer}
\end{figure*}

SGR$\,$1900$+$14 is one of the youngest magnetars known ($\sim$1\,kyr).  A very prolific burster \citep{Israel08}, it is embedded in a cluster of very massive stars \citep{Vrba00, Davies09} and it is one of the three magnetars which has shown a Giant Flare. This magnetar was observed using all three instruments onboard the NASA {\it Spitzer Space Telescope} in 2005 and 2007 \citep{Wachter08}.
Surprisingly, these observations have revealed a prominent ring-like structure (see Figure\,1) in the 16$\mu$m and 24$\mu$m wave-bands, not detected in the 3.6--8.0\,$\mu$m observations. A formal elliptical fit to the ring indicates semi-major and semi-minor axes of angular lengths $\sim\,$36'' and $\sim\,$19'', respectively, centred at the position of the magnetar SGR$\,$1900$+$14. No equivalent feature was observed at radio or X-ray wavelengths ($L_{\rm 332 MHz} \leq 2.7 \times 10^{29}\;d_{12.5}^2\;{\rm erg\;s^{-1}}$, and $L_{\rm 2-10 keV} \leq 1.8 \times 10^{33}\;d_{12.5}^2\;{\rm erg\;s^{-1}}$; with $d_{12.5}$ being the distance in units of 12.5\,kpc; \citet{Kaplan03}).

The {\it Spitzer} images are dominated by the bright emission from two nearby M supergiants that mark the centre of a compact cluster of massive stars at a distance of $\sim$12.5\,kpc, believed to have hosted the magnetar progenitor star \citep{Vrba96, Vrba00, Davies09}. The physical size of the ring at this distance is $\sim 2.18\times1.15$\,pc, it has a temperature $\sim 80$\,K \citep[estimated using a realistic dust model:][]{Draine03, Wachter08}, and a flux of 0.4$\pm$0.1\,Jy and 1.2$\pm$0.2\,Jy at 16 and 24$\mu$m, respectively \citep[corresponding to L(16$\mu$m)$\sim1.4\times10^{36}d^2_{12.5}$\ergs and L(24$\mu$m)$\sim2.7\times10^{36}d^2_{12.5}$\ergs][]{Wachter08}. However, the inevitable difficulties in the analysis of this faint and complicated structure make these values rather uncertain and preclude a detailed assessment of the true ring morphology. The ring-like structure has been interpreted as due to illumination from nearby stars of a dust free cavity produced by the Giant Flare \citep{Wachter08}.

In this work we show the results of a series of radiation transfer (RT) calculations performed with a 3D dust radiative transfer code {\em DART-Ray}  \citep{Natale14, Natale15} aimed at reproducing the emission of the infrared ring around SGR$\,$1900$+$14, assuming different plausible distributions for the dust illuminated by the nearby stars. In \S\ref{setup} we introduce our initial assumptions, method, and the RT calculations. Results and Discussion follow in \S\ref{results} and \S\ref{discussion}.

\begin{table}[t]
\caption{Magnetar and close-by star properties$^{*}$}              
\label{table_star_positions}      
\centering                                      
\begin{tabular}{c c c c c }          
\hline\hline                        
Star & R.A. & Dec & T$_{\rm eff}$ & log (L$_{\rm bol}$/L$_{\odot}$) \\    
    & [h min sec] & [grad min sec]  &  K  &  \\
\hline                                   
  1900+14 & 19 07 14.33 &  09 19 20 \\ 
   A (M2)  &  19 07 15.34 & 09 19 21.7 &  3660 & 5.05   \\
   B (M1)  & 19 07 15.12 & 09 19 21.0  & 3750 & 4.91   \\ 
\hline                                             
\end{tabular}
\begin{list}{}{}
\item[$^{*}$] {\it As derived by \citet{Davies09}.}
\end{list}
\end{table}

\section{Radiation transfer calculations}
\label{setup}

In order to set up radiation transfer models appropriate to reproduce the observed ring emission, we first defined the distribution of stars and dust within the volume over which the RT calculations are performed (a cube of size 10 pc). In this section, we describe how this has been done given the constraints provided by the observations. Specifically, in \S\ref{dust_ring_properties} the assumed ring distance, physical sizes and fluxes are given. In \S\ref{section_geom_descr_ell} we explain how the dust distributions have been defined. In \S\ref{section_lines_of_sight} we describe how we found the appropriate viewing angles for the observer such that the projected ellipse is approximately of the same size and orientation as the dust emission ring. In \S\ref{section_star_3d_pos} we show how we derived the intrinsic 3D positions of the stars, relative to the magnetar, by using the constraint given by their projected positions on the sky. Finally, in \S\ref{simulations} we describe the specific RT models we considered and how the RT and dust emission calculations have been performed.    

\subsection{Ring distance, sizes and mid-infrared fluxes}
\label{dust_ring_properties}

We set up the size and geometry of the assumed 3D dust distributions by comparison with the observations (see Figure\,\ref{spitzer}). We considered the distance measurement of \citet{Davies09}, which found d=12.5 $\pm$1.4\,kpc by using optical spectroscopy of the close-by stars. Assuming d=12.5\,kpc, the lengths of the projected ring major and minor semi-axis are 2.18 and 1.15 pc. The major axis of the ring is rotated by about 22\,\deg from the R.A. axis. We considered the integrated fluxes for the dust ring in the mid-infrared as measured by \citep{Wachter08}: F(24$\mu$m)=1.2$\pm$0.2\,Jy and F(16$\mu$m)= 0.4$\pm$0.1\,Jy.

\begin{figure*}[t]
\includegraphics[scale=0.5]{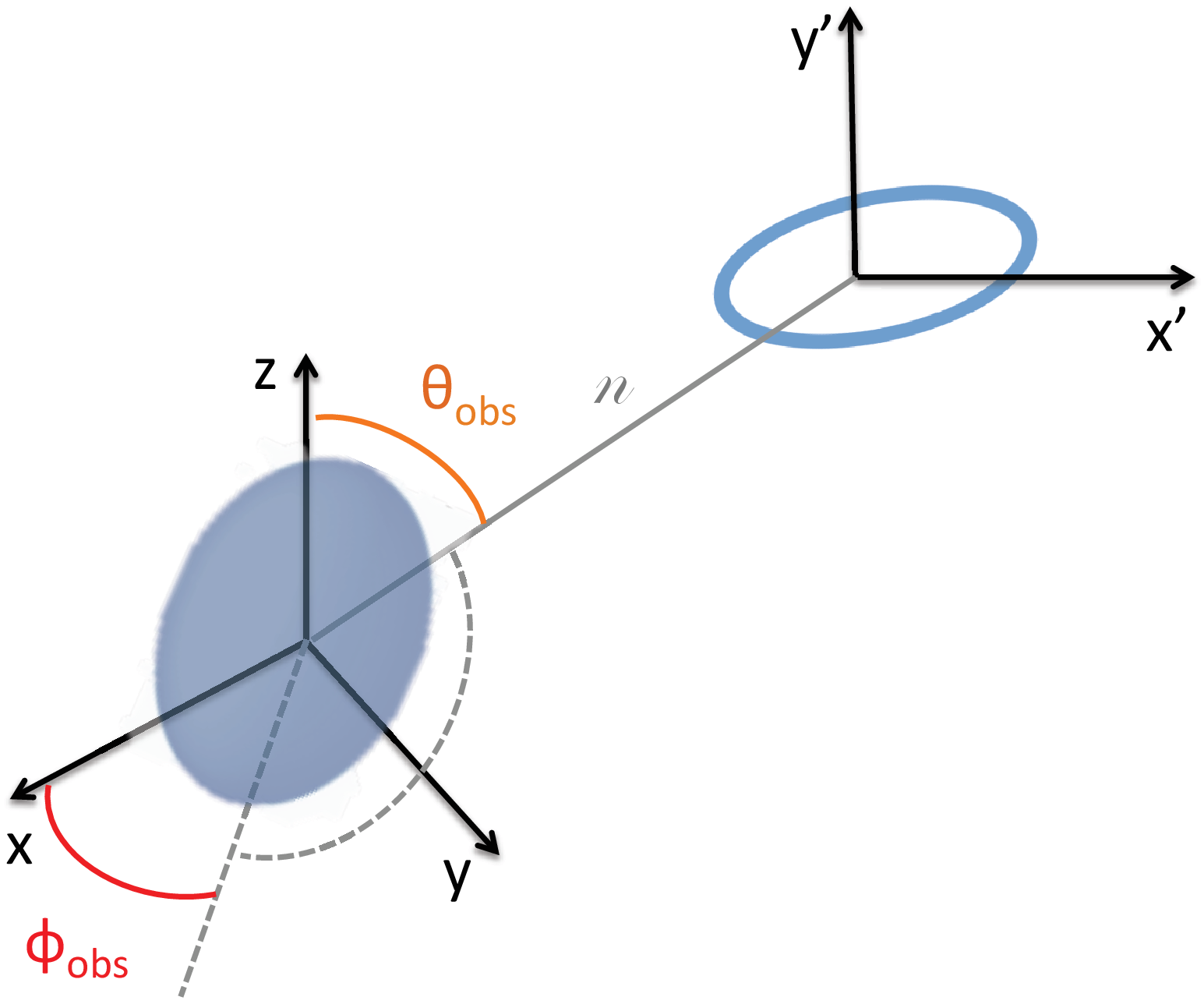}
\includegraphics[scale=0.5, trim={3.5cm 5cm 0 0},clip]{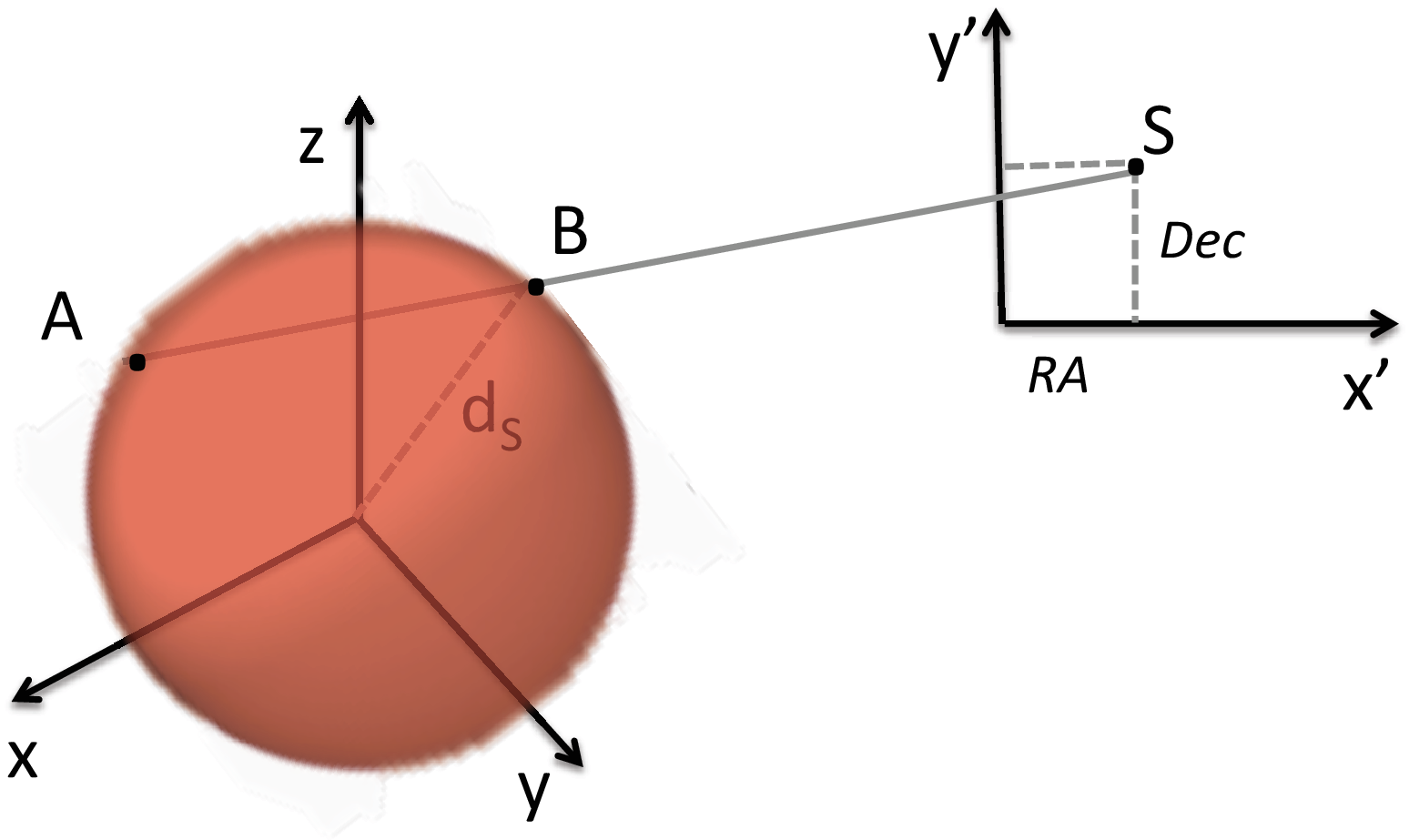}
\caption{{\em Left panel}: Sketch of the reference frames involved in the ellipsoid projection. The frame "xyz" is the reference frame where the ellipsoid is defined. The 2D frame x'y' defines the observer plane, over which the ellipsoid is projected. The line "n" is the observer line-of-sight which is perpendicular to the observer plane and inclined by the angles $\theta_{\rm obs}$ and $\phi_{\rm obs}$ with respect to the xyz reference frame. {\em Right panel}: Sketch showing the method to find plausible positions for the stars within the 3D reference frame xyz. For a given R.A. and Dec, a star can be located only on the line perpendicular to the observer plane x'y' and intersecting it in the point $S$ with sky coordinates (R.A., Dec). Then, we parametrize the 3D position of the star in terms of the distance $d_s$ from the magnetar (located in the origin of the 3D reference frame xyz). Geometrically, for a given $d_s$, the two possible points on the line where the star can be located correspond to the intersection of the line with a sphere of radius $d_s$ centered on the magnetar.  }
\label{fig_sketch_ref}
\label{fig_find_pos_stars}
\end{figure*}


\subsection{Assumed ellipsoidal dust distributions}
\label{section_geom_descr_ell}
The observed 2D ring-like emission morphology observed on the {\em Spitzer} data is compatible with being the projection of a 3D ellipsoidal structure. We modelled it with a thin ellipsoidal shell, a uniform dust distribution around an ellipsoidal cavity or a more complicated distribution resulting from a stellar wind dust density profile which has been internally depleted of dust. In this section, we show how we have defined dust density profiles representative of each of these cases. The three profiles we have chosen should be considered as simplified representations of the complex dust distributions determined by the physical processes plausibly giving rise to the 2D ring we observe. Given the large uncertainties on the details of these physical processes, we could only assume simple shapes which qualitatively reproduce the expected dust distributions. 

The ellipsoidal surfaces, needed to define the above ellipsoidal structures, are described by the following formula for a constant normalized radius $R$:

\begin{equation}
\label{normal_radius}
R^2=\frac{x^2}{a^2}+\frac{y^2}{b^2}+\frac{z^2}{c^2}
\end{equation}

where $a$, $b$ and $c$ are the lengths of the three semi-axis of the "reference ellipsoid" with $R=1$. If for a given (x,y,z) we have $R>1$ or $< 1$, the point (x,y,z) belongs to an ellipsoidal surface which is either inside or outside the reference ellipsoid (but it has the same axis ratio). In order to consider the volume within an ellipsoidal shell, that is, the volume embedded between two ellipsoidal surfaces with different normalized radii, we consider all the points (x,y,z) satisfying the following relation:

\begin{equation}
\sqrt{| R^2 -1| } < \Delta R
\label{formula_ellipoid_shell}
\end{equation}   

where $\Delta R$ represents the semi-width of the ellipsoidal shell (in normalized $R$ units). For the elliptical shell distribution, the dust density is assumed to be constant within the volume defined by Eq.\ref{formula_ellipoid_shell} and zero outside. For the ellipsoidal cavity distribution, we assumed that the dust density is constant for $R>1$ and zero for $R<1$. Finally, for the stellar wind distribution, we assumed the following dust density radial profile: 
\begin{align*}
\rho_d(R)=\rho_d(R_d)\left(\frac{R}{R_d}\right)^{2}     &~~    \rm{if} ~R<R_d, \\
\rho_d(R)=\rho_d(R_d)\left(\frac{R}{R_d}\right)^{-2}     &~~    \rm{if} ~R>R_d 
\end{align*} 
with $R_d=1$. The wind density profile for $R>1$ decreases as $R^{-2}$ and thus resembles that expected in a stellar wind with elliptical symmetry. The profile for $R<1$ is hard to predict theoretically, since it is due to dust destruction processes with unknown parameters. For the sake of simplicity, we assumed it rises as $R^2$ until R=1. The three types of density profiles we defined are shown in Figure\,\ref{fig_magtar_dust_profiles}.

\subsection{Derivation of the lines-of-sight reproducing the ring apparent sizes}
\label{section_lines_of_sight}
For all the above dust distribution profiles,  the axis lengths and orientation of the ring are determined by the intrinsic lengths of the ellipsoid axis as well as by three angles: two angles ($\theta_{\rm obs}$ and $\phi_{\rm obs}$) which specify the line-of-sight direction (see left panel of Fig. \ref{fig_sketch_ref}), and one angle that specifies the orientation of the 2D reference frame on the observer plane. By "observer plane", we mean the plane over which the ellipsoid is projected (that is, simply the plane of the data map). The ellipsoid projection is degenerate in the sense that different combinations of the ellipsoid intrinsic parameters can produce the same projected ellipse just by choosing appropriate observer viewing angles.

In order to be able to predict the shape of the projected ellipse for arbitrary combinations of the ellipsoid and observer parameters, we have used the formulae derived by \citet[GS81, see section "Plane projections of a triaxial ellipsoid", equations 25]{Gendzwill81}. Given an ellipsoid and the observer plane, these authors derived analytical formulae for the projected ellipse by determining all the points on the observer plane where the normal to the plane is tangent to the three-dimensional ellipsoid. By using equations 25 of GS81, we are able to predict the projected ellipse semi-axis and orientation given the ellipsoid semi-axis $a$, $b$, $c$ and the observer line of sight angles $\theta_{\rm obs}$ and $\phi_{\rm obs}$\footnote{The formulae in GS81 are written in terms of three rotation angles $\theta_{GS}$, $\phi_{GS}$ and $\psi_{GS}$ (see Fig. 3 in GS81) which can be connected to our definition of observer viewing angles $\theta_{\rm obs}$ and $\phi_{\rm obs}$ in the following way:
\begin{equation}
\theta_{GS}=\frac{\pi}{2} \quad
\phi_{GS}=\frac{\pi}{2}-\theta_{\rm obs} \quad 
\psi_{GS}=\phi_{\rm obs}-\frac{\pi}{2}.
\end{equation}
By using the first relation, we force one axis of the 2D reference frame on the observer plane to be the projection of the $z$-axis of the 3D frame "xyz", where the ellipsoid is defined. In this way, $\theta_{\rm obs}$ and $\phi_{obs}$ are easily related to the angles $\phi_{GS}$ and $\psi_{GS}$ in GS81 using the second and third relation above.}. 
Then, in order to handle the inverse problem of finding the combinations of the observer angles $\theta_{\rm obs}$ and $\phi_{\rm obs}$ that produce a projected ellipse with the same parameters of the observed dust emission ring, we wrote an optimization program which finds the right $\theta_{\rm obs}$ and $\phi_{\rm obs}$ for a given combination of ellipsoid semi-axis $a$, $b$, and $c$. Specifically, we choose to fix the values of $b$ and $c$ to 2.18 pc, the length of semi-major axis of the projected ellipse. We then assumed $b/a$= 2 or 4. For these different combinations of ellipsoid parameters, we found the values of $\theta_{\rm obs}$ and $\phi_{\rm obs}$ which allow to reproduce the shape and orientation of the projected ellipse (see \S\ref{dust_ring_properties}). The values for $\theta_{\rm obs}$ and $\phi_{\rm obs}$ we derived are listed in Table \ref{table_obs_angles}. 

\begin{table}
\caption{Observers angle$^*$}              
\label{table_obs_angles}      
\centering                                      
\begin{tabular}{c c c}          
\hline\hline                        
$b/a$ & $\theta_{\rm obs}$ & $\phi_{\rm obs}$ \\    
    & [deg] & [deg] \\
\hline                                   
   2 & 12.05 & 21.56 \\
   4 & 30.52 & 19.19 \\
\hline      
\end{tabular}                                       
\begin{list}{}{}
\item[$^{*}$] {\it Derived observer angles $\theta_{\rm obs}$ and $\phi_{\rm obs}$, depending on assumed $b/a$ axis ratio and assuming $b=c=2.18$\,pc.}
\end{list}
\end{table}

\subsection{Derivation of the intrinsic 3D positions of the stars from their sky position}
\label{section_star_3d_pos}

In all the RT models that we calculated, we placed the magnetar at the origin of the 3D reference frame xyz, since the magnetar appears at the centre of the projected ellipse. For the two supergiant stars we had to find a way to place them such that their projected position on the observer plane coincides with their sky coordinates R.A. and Dec (see Table \ref{table_star_positions}). 
Because this is the only constraint we have, each star can in principle be located at any point on a line perpendicular to the observer plane and intersecting the observer plane in (R.A., Dec), as shown in the right panel of Fig.\ref{fig_find_pos_stars}. We parametrize the position of a star along this line in terms of its distance $d_s$ from the magnetar. As shown in Fig.\ref{fig_find_pos_stars}, for each value of $d_s$ there can be up to two possible positions for the star (but also just one or even none if $d_s$ is too small). If two possible positions exist, one of the two is chosen as location for the star. In Table\,\ref{table_star_3d_pos} we show the positions we derived for the stars assuming they are located at different distances $d_s$ from the magnetar (that is, $d_s$ = 1 and 3 pc). These distances are chosen such that the stars are either within the ellipsoidal cavity, or outside it but not too far from its border. We call these two types of star locations "IN" and "OUT" configurations.

\begin{figure}
\includegraphics[scale=0.5]{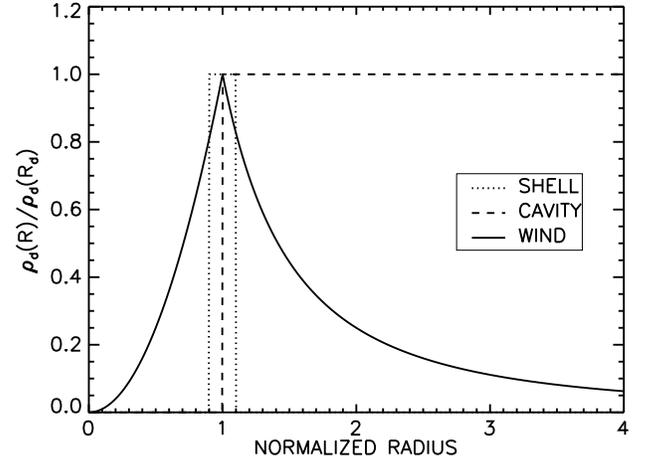}
\caption{The three types of density profiles used to model the ring emission.}
\label{fig_magtar_dust_profiles}
\end{figure}


\begin{table}[t]
\caption{3D positions of the stars$^{*}$.}              
\label{table_star_3d_pos}      
\centering                                      
\begin{tabular}{l| c c c c c c c}          
\hline\hline                        
Geometry & Star & $b/a$ &  $d_s$ & $R$ & x & y & z \\    
 &   &  & [pc] & & [pc] & [pc] & [pc] \\
\hline                                   
IN& A  &   2 &  1 & 0.61 & -0.50 & 0.79 & -0.35 \\  
 & B  &   2 &  1 & 0.58 & -0.45 & 0.59 & -0.67 \\
\hline 
IN& A &   4 &  1 & 0.58 & -0.19 & 0.90 & 0.38 \\
 & B  &   4 &  1 & 0.47 &  0.05 & 0.78 & 0.62 \\
\hline                                   
OUT & A  &   2 &  3 & 1.58 & -0.98 & 0.60 & -2.77 \\
 & B   &   2 &  3 & 1.54 & -0.88 & 0.42 & -2.84 \\
\hline                                             
\end{tabular}
\begin{list}{}{}
\item[$^{*}$] {\it Derived 3D stellar positions (x,y,z) depending on the assumed $b/a$ axis ratio and the assumed distance $d_s$ of the stars from the magnetar located in the centre of the Cartesian reference frame.  For each stellar position, we also show the normalized radius value $R$ which indicates the position of the stars relative to the reference ellipsoid. }
\end{list}{}{}
\end{table}

\subsection{3D dust radiative transfer and dust emission calculations}
\label{simulations}

In order to calculate the dust heating from the stars and the resulting dust emission, we used the 3D ray-tracing dust radiative transfer code {\em DART-Ray}   \citep{Natale14, Natale15}. This code allows to solve the 3D dust radiative transfer problem for arbitrary distributions of dust and stars \citep[see review of][]{ Steinacker13}: it follows the propagation of the light emitted by the stars within an RT model, including absorption and multiple scattering due to the dust. By calculating the variation of the light specific intensity $I_\lambda$ throughout a RT model, it also derives the distribution of the radiation field energy density $u_\lambda(\vec{x})$ of the UV/optical/near-infrared radiation:
\begin{equation}
u_\lambda(\vec{x})=\frac{\int I_\lambda(\vec{x},\vec{n})d\Omega}{c}
\end{equation}
 From $u_\lambda$ and the dust density distribution, the dust emission can then be calculated at each position. We performed the dust emission calculation taking into account the stochastic heating of small grains \citep[see e.g.][]{Draine03}. This is important to consider since small grains tend to not achieve equilibrium with the interstellar radiation field and, heated by single photons carrying energies comparable to their internal energies, experience large temperature fluctuations. Thus, taking into account stochatical heating is important to correctly derive the dust emission spectra in particular in the mid-infrared. Unlike the equilibrium case, where a grain is characterized by only one dust temperature, in the stochastically heated case a grain has a certain probability P(T) to be at a certain temperature T. At each position inside the RT model, the probability function P(T) is calculated for each grain size $a$ and composition $k$ following \citet{Voit91} \citep[see also][] {Natale15}. 
The function P(T) depends on both the local value of $u_\lambda$ and the grain absorption coefficient $Q^{\rm abs}_\lambda(a,k)$. Once P(T) is derived, the dust emission luminosity for a single grain $\epsilon_\lambda$(a,k) can then be calculated as:
\begin{equation}
\epsilon_\lambda(a,k)=4\pi^2 a^2Q^{\rm abs}_\lambda(a,k)\int_0^{\infty} P(T)B_\lambda(T)dT
\end{equation} 
where $B_\lambda$ is the Planck function. The total dust emission at each position is derived after integration over the grain size distribution (for each chemical species). 
Finally, by projecting the dust emission at each position onto the observer plane and by convolving the resulting maps to the instrument angular resolution (FWHM$\sim$5.3 and 6\,arcs for Spitzer 16 and 24\,$\mu$m), dust emission maps can be derived as they would be obtained by an external observer. 

As input, {\em DART-Ray}  just needs a 3D Cartesian adaptive grid where for each element the dust density and the stellar volume emissivity are specified. Both distributed stellar emission and stellar point sources can be included. We created input grids representing the three types of dust distributions described in \S\ref{section_geom_descr_ell}: an ellipsoidal shell, an ellipsoidal cavity and a ellipsoidal wind.
As we already mentioned in \S\ref{section_star_3d_pos}, we assumed  $b=c=2.18$\,pc and $b/a$=2 or 4. In the case of the ellipsoidal shell dust distribution, we also assumed $\Delta R=0.10$. We chose this value for $\Delta R$ because it produces a dust emission ring with approximately the same width of the observed ring. 

To set the dust optical properties (absorption and scattering efficiency, scattering phase angle parameter, dust-to-gas mass ratio), we assumed the dust model BARE-GR-S of \citet{Zubko04} as implemented in the TRUST dust RT benchmark project \citep{Camps15}. In this dust model, the dust is composed by a mix of silicates, graphite and PAH molecules \citep[see Fig. 4 of][showing the size distribution of each component]{Zubko04}. The abundance of each component in the dust model has been chosen to match observational data for the average extinction curves, chemical abundances and dust emission in the Milky Way.  

To set the dust/gas density in the RT model, we assumed several choices for the optical depth per unit parsec at 1\,$\mu$m which, assuming the dust model mentioned above, correspond to a wide range of values for the hydrogen number density $n_H$. Since the density can vary within an RT model, the density values we will mention hereafter refer to the density at the reference ellipsoid (that is, for R = 1). Because of the long time required for a single RT/dust emission calculation (several hours), the density values have not been varied automatically in order to minimize the disagreement with the data but have been simply chosen in order to represent a typical diffuse interstellar medium (ISM) density ($n_H$=10\,cm$^{-3}$) as well as much denser medium ($n_H$=100 and 1000\,cm$^{-3}$, see Table \ref{table_tot_mid_fluxes}). The ring integrated fluxes for the models with $n_H$=1000\,cm$^{-3}$ are compatible with the observed fluxes, as it will be shown in the next section.

Following the procedure described in \S\ref{section_star_3d_pos}, we positioned the two M supergiant stars at either $d_s$=1 or $d_s=3$\,pc from the magnetar (the two stars are at same distance from the magnetar but at two different 3D positions since they are projected in two different points on the observer plane). We assume these stars are emitting as blackbodies with effective temperatures and bolometric luminosities as found by \citet{Davies09} (see Table \ref{table_star_positions}). 

\begin{table*}
\centering
\small
\caption{RT Model parameters and mid-infrared ring emission fluxes.}
\label{table_tot_mid_fluxes}                            
\begin{tabular}{c c c c c c}          
\hline
\hline      
\multicolumn{6}{c}{Ellipsoidal shell}\\
\hline                  
STARS & $\tau_{1\mu m}$/(1 pc) & $n_H$ &  $b/a$ &  F(16$\mu$m) & F(24$\mu$m) \\    
  &   & [$cm^{-3}$]  &  & [Jy] & [Jy] \\
\hline                                   
IN &  4.8$\times10^{-3} $ & 10  & 2  &  0.0031 & 0.011 \\
OUT & 4.8$\times10^{-3} $ & 10  & 2  &   3.2E-4 & 8.1E-4 \\
IN &  4.8$\times10^{-3} $ & 10  &  4  &   0.0014 & 0.0049 \\
IN &  4.8$\times10^{-2} $ & 100  & 2  &   0.024 & 0.056 \\
IN &  4.8$\times10^{-1} $ & 1000  & 2 &   0.28 & 0.98 \\ 
\hline                     
\multicolumn{6}{c}{Ellipsoidal cavity}\\
\hline
IN &  4.8$\times10^{-3} $ & 10  & 2   &  0.0027 & 0.011 \\
OUT &  4.8$\times10^{-3} $ & 10  & 2   & -- & -- \\
IN &  4.8$\times10^{-2} $ & 100  & 2   & 0.028 & 0.11 \\ 
IN &  4.8$\times10^{-1} $ & 1000  & 2  & 0.36 & 1.35 \\
\hline                                
 \multicolumn{6}{c}{Ellipsoidal wind}\\
\hline        
IN &  4.8$\times10^{-1} $ & 1000  & 2  & 0.28 & 1.06 \\
OUT &  4.8$\times10^{-1} $ & 1000  & 2  & -- & -- \\
\hline
\end{tabular}
\end{table*}

\section{Results}
\label{results}

\begin{figure*}[t]
\centering
\includegraphics[scale=0.9, trim = 0 0.5cm 0 0]{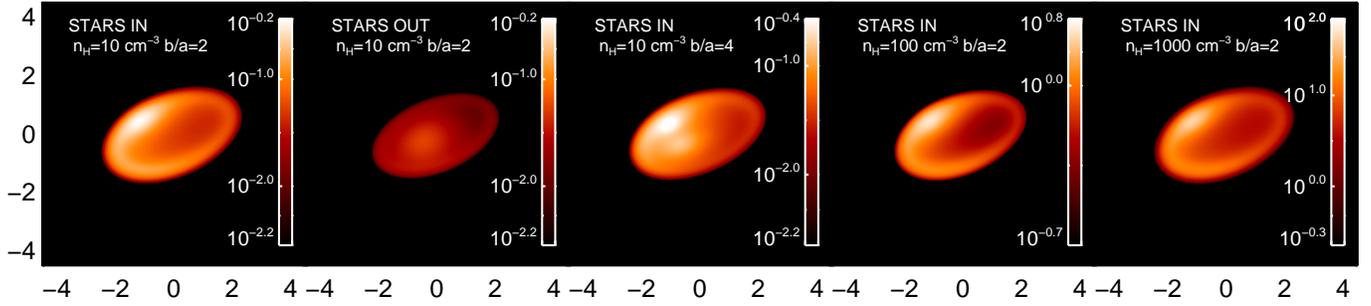}
\vspace{0.cm}
\caption{Synthetic dust emission maps at 16$\mu$m for the ellipsoidal shell models. The position of the stars is inside the dust cavity for the "IN" models and outside for the "OUT" models. The dust emission is calculated by taking into account the stochastical heating of the dust (see text). Units on the color bar are [MJy/sr]. The figure axis show the projected distance from the magnetar in pc. } 
\label{fig_magtar_16um_shell}
\end{figure*}

\begin{figure*}[t]
\centering
\includegraphics[scale=0.9]{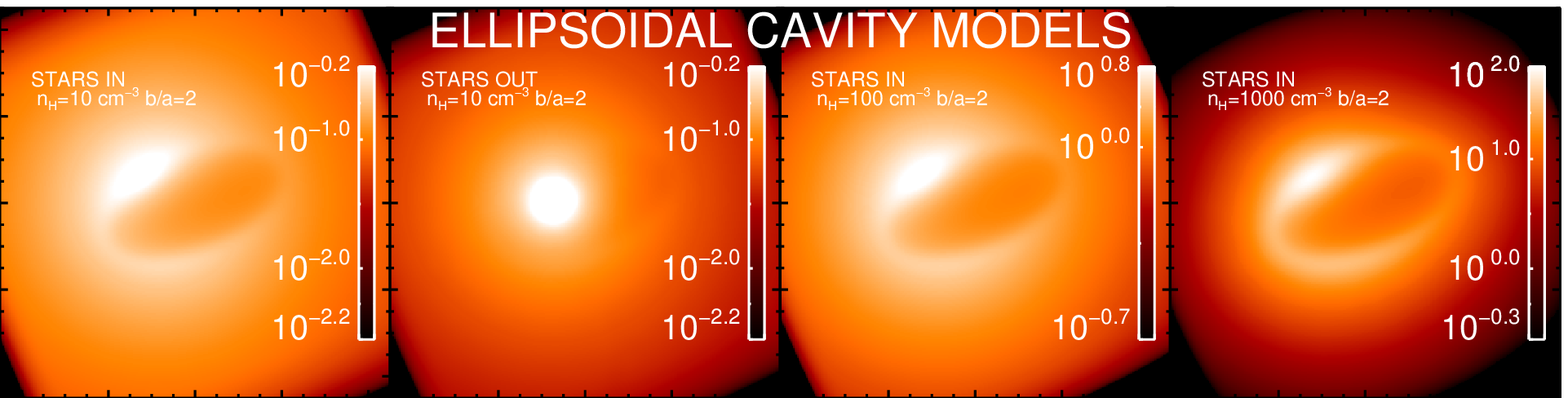}
\vspace{0.5cm}
\caption{Synthetic dust emission maps at 16$\mu$m for the ellipsoidal cavity model. Same description as in Fig. \ref{fig_magtar_16um_shell}. }  
\label{fig_magtar_16um_cavity}
\end{figure*}

\begin{figure*}[t]
\centering
\includegraphics[scale=0.5]{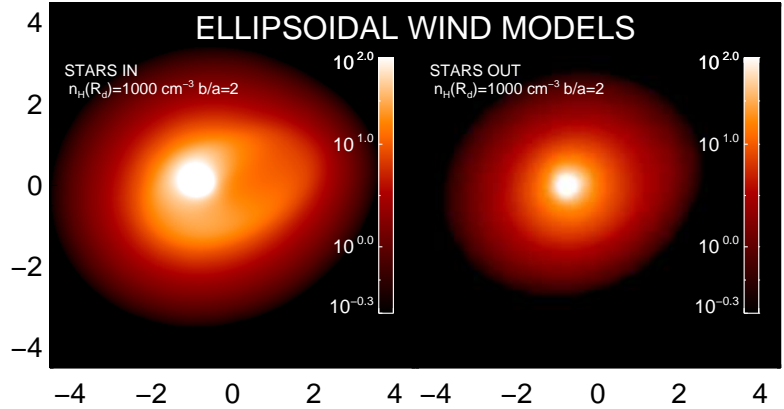}
\vspace{0.5cm}
\caption{Synthetic dust emission maps at 16$\mu$m for the ellipsoidal wind model. Same description as in Fig. \ref{fig_magtar_16um_shell}. } 
\label{fig_magtar_16um_wind}
\end{figure*}

In this section we show the results we obtained for the various assumed dust ellipsoidal distributions (see Figure\, \ref{fig_magtar_dust_profiles}), dust densities (defined by the 1\,$\mu$m optical depth per unit length or, equivalently, the hydrogen number density $n_H$), the stellar positions (which can be in the IN or OUT configurations) and the reference ellipsoid axis-ratio b/a. 
The parameters of the calculated RT models are shown in Table \ref{table_tot_mid_fluxes}. In the same table we also show the fluxes we measured for the dust emission ring appearing on the 16 and 24 $\mu$m model maps. The ring photometry has been taken on an elliptical aperture containing the ring, in analogy to the measurement of \citep{Wachter08} on the {\em Spitzer} maps. The 16\,$\mu$m emission model maps, corresponding to all the RT models listed in Table \ref{table_tot_mid_fluxes}, are shown in Figures  \ref{fig_magtar_16um_shell}, \ref{fig_magtar_16um_cavity} and \ref{fig_magtar_16um_wind}. The variation of the parameters listed above has a clear effect on the morphology of the synthetic maps and/or on the total flux. 

\subsection{Ellipsoidal shell models}

The first two panels on the left in figure \ref{fig_magtar_16um_shell} show the maps obtained for the ellipsoidal shell RT model with $n_H=10~$cm$^{-3}$ and $b/a=2$ for the IN and OUT configurations.  As one can see, only the IN configuration gives rise to a clear dust ring shape, while the OUT configuration presents an additional bright feature within the ring--like shape. Although this enhancement resembles the one seen on the Spitzer maps, its nature is very different. The enhancement on the data maps appears at the position of the supergiant stars, it is a PSF-convolved point source and it is much brighter than the emission from the ring. Being much brighter than the expected blackbody emission for those stars, this emission is probably due to dust in the circumstellar material around the M supergiants. On the other hand, the bright feature observed in the OUT models is extended and shifted with respect to the position of the stars. This emission is due to dust in the elliptical shell that is closer to the stars located outside the shell and, thus, heated more strongly by them. 

The effect of modifying the b/a ratio, but keeping the stars inside the cavity, can be seen from the middle panel. Also in this case, although a dust ring is visible, an emission enhancement with comparable brightness to that of the ring is visible on the map. This is due to the more elongated shape of the cavity where some parts of the internal boundary of the cavity are  significantly more heated by the stars, thus causing this effect. In terms of emission flux, in this model the integrated ring flux is about a half of that of the model with $b/a=2$ (although the flux integrated over the entire map is higher). More importantly, the change of total flux due to the variation of $b/a$ is too small to allow to reproduce the observed fluxes while keeping $n_H=10~$cm$^{-3}$. This is because the $b/a$ ratio can be varied, realistically, only within a rather small range of values ($\approx 1-10$).   

On the other hand, the dust density can in principle be varied over a wide range of values and, thus, can give rise to a substantial change in the total flux without affecting the emission morphology. We performed calculations with dust densities corresponding to $n_H$=100 and 1000\,cm$^{-3}$ for a model with b/a=2 and stars within the cavity. As one can see from the right panels of Figure \ref{fig_magtar_16um_shell} and the ring fluxes in Table \ref{table_tot_mid_fluxes}, the predicted emission has the same morphology of the $n_H$=10\,cm$^{-3}$ model but it is much more luminous. The dust emission SEDs of the ring for these two models are shown in Fig. \ref{fig_magtar_sed} (left panel). This plot shows that for the model with $n_H$=1000\,cm$^{-3}$ the fluxes we obtained are within 1.5$\sigma$ from those measured by \citep{Wachter08}.

\subsection{Ellipsoidal cavity models}

We also performed RT calculations assuming a dust cavity geometry, where the dust is uniformly distributed outside the reference ellipsoid. This is expected in case the flare from the magnetar simply caused dust destruction within the dust cavity and did not affect significantly the density of the surrounding medium (thus, there is no significant enhancement of the gas/dust density close to the border of the cavity in this case). For $n_H$=10\,cm$^{-3}$ and b/a =2, we calculated the maps for the IN and OUT geometry. The corresponding 16$\mu$m maps are shown in Figure \ref{fig_magtar_16um_cavity}. In both cases the emission is much more extended compared to that of the shell models (which do not contain any dust apart from that within the shell). However, the ring emission morphology is clearly recovered only in the case where the stars are inside the cavity. If the stars are outside, they illuminate the dust close to them very strongly. The emission from this dust dominates the emission seen on the maps. Instead, the ring--like emission is barely noticeable, with brightness close to that of the background emission. We point out that the bright dust emission seen in this case is much more extended than the point source emission seen on the data map at the position of the stars. We also calculated higher density models ($n_H$=100-1000\,cm$^{-3}$) for the IN configuration. The maps shown in Fig.\ref{fig_magtar_16um_cavity} are characterized by a higher surface brightness, compared to the calculation with $n_H$=10\,cm$^{-3}$, but a similar emission morphology. 

The ring fluxes for the ellipsoidal cavity model and for the IN configuration are listed in Table \ref{table_tot_mid_fluxes}. These can be compared with the ring fluxes for the corresponding shell models with same parameters. As one can see, there are some differences between the shell and cavity models with same parameters, but the fluxes are similar. We also show the ring dust emission SEDs for the highest density models in Fig. \ref{fig_magtar_sed} (middle panel).  The cavity model with  $n_H$=1000\,cm$^{-3}$ fits the observed integrated dust emission within the data error bars. This shows that both a thin dust shell and a dust cavity RT model give similar results for the ring integrated fluxes and morphology in the case that the stars are located inside the cavity. Also, overall, the results we obtained from the dust cavity models further suggest that the presence of the stars (or other radiation sources) within the cavity is a necessary requirement to recover the observed dust emission morphology. 

\subsection{Ellipsoidal dusty wind models}

The last geometry we explored for the dust is that expected in the case the dust around the magnetar is distributed as in a stellar wind (with elliptical symmetry) and the dust has been only partially destroyed within the cavity region. The assumed density profile rises until the border of the cavity and then decreases as $R^{-2}$ as shown in Sect.\ref{section_geom_descr_ell}. Given the previous results for the shell and cavity model, we have run only RT calculations assuming $n_H=1000~\rm{cm^{-3}}$ at $R=1$ for this model. The results are shown in Fig. \ref{fig_magtar_16um_wind} for both the IN and OUT configurations for the stars. In the case the stars are located inside, the morphology of the ring is recovered although not as clearly as we found before for some of the shell and cavity models.  There is substantial diffuse emission coming from the regions enclosed by the ring. Interestingly, a quite narrow and bright dust emission appears at the position of the stars. This feature resembles the bright dust emission source seen on the data at the star positions. A sharper profile, rising more rapidly close to $R=1$, would certainly reduce the brightness of this diffuse emission, as observed for the two previous configurations. However, given the large uncertainties on dust grain sizes or the properties of the destructing flare, we assume this dust density shape as a toy model to test the wind scenario.

\begin{figure*}
\centering
\includegraphics[scale=0.35]{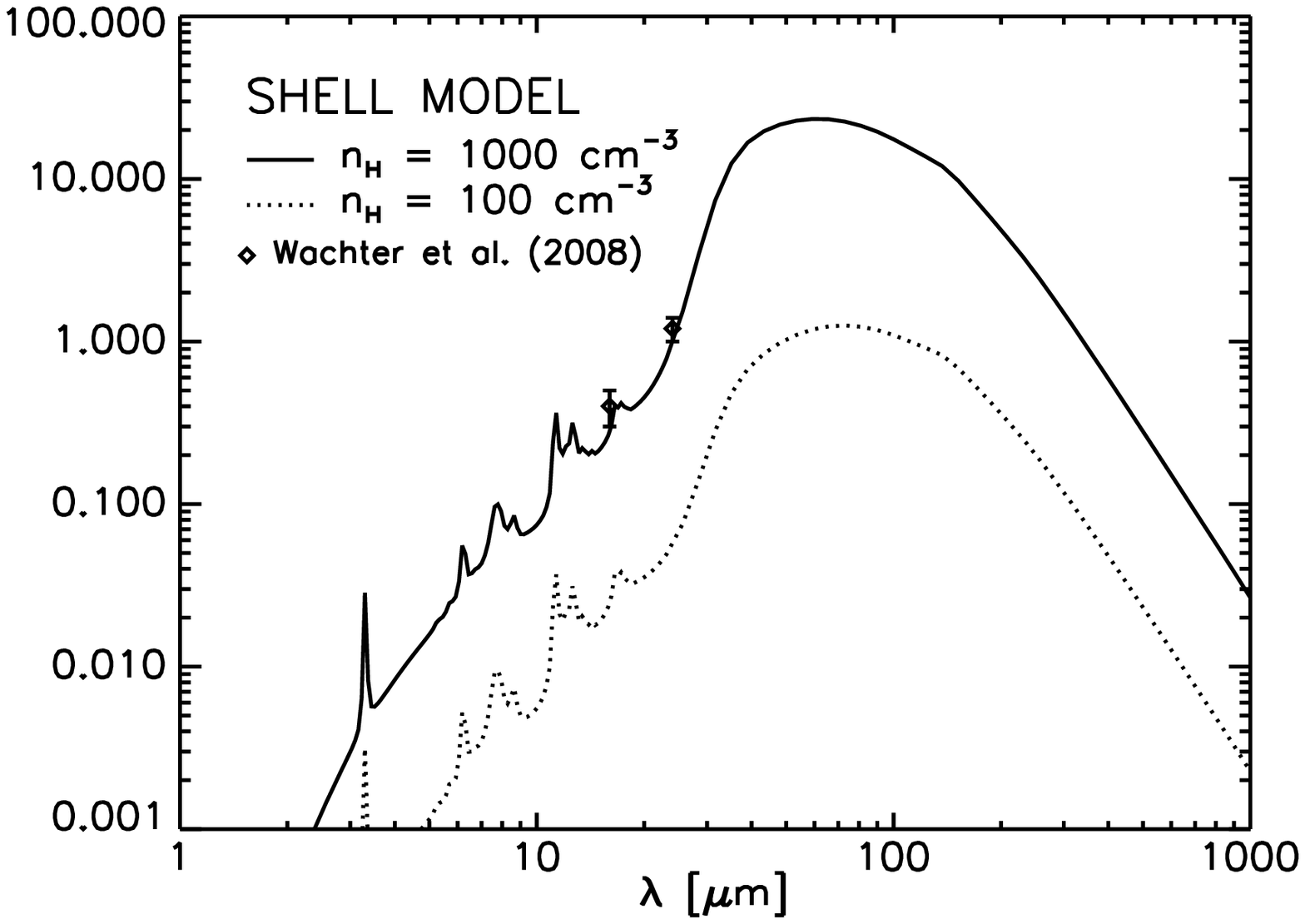}
\includegraphics[scale=0.35, trim={2.5cm 0 0 0},clip]{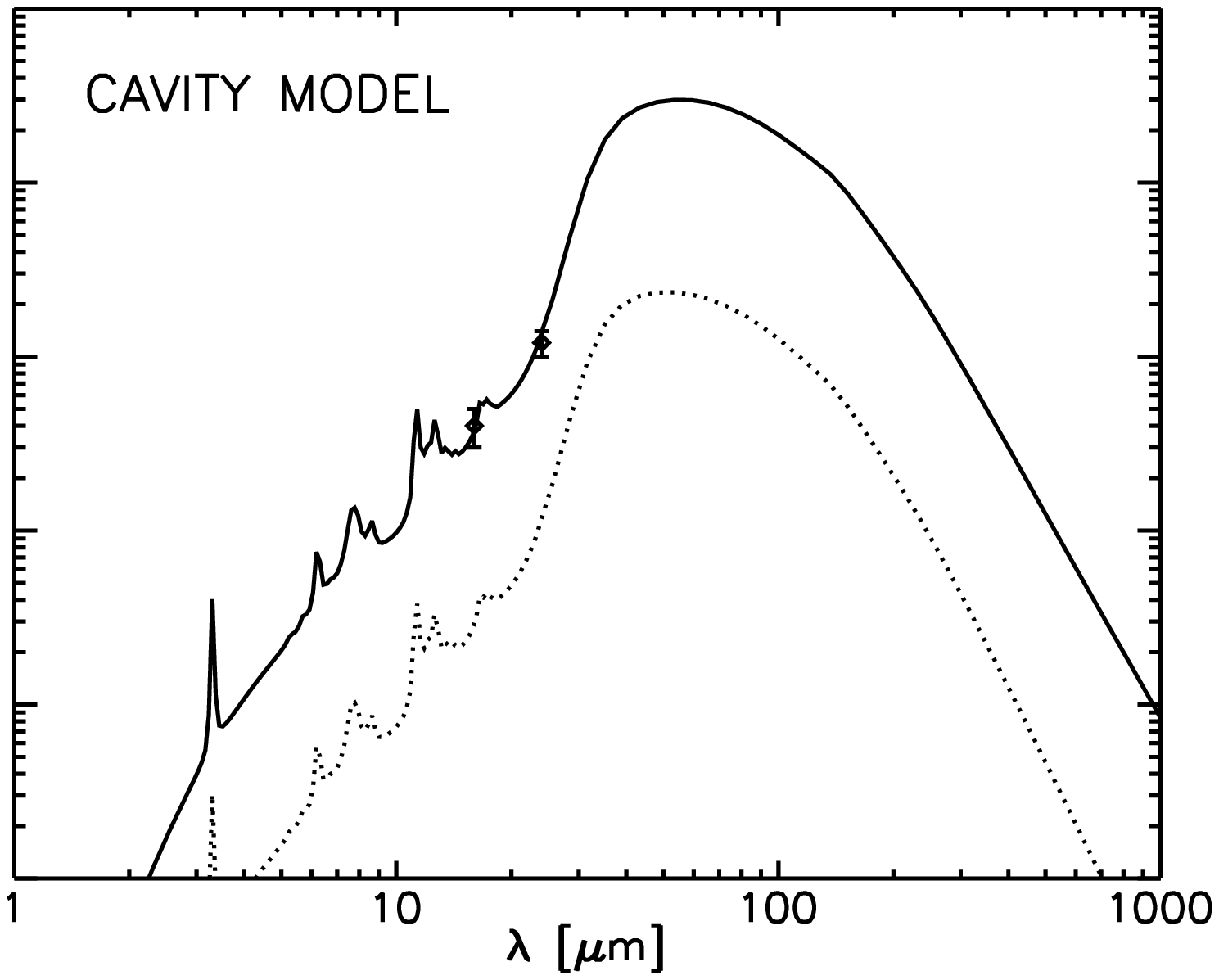}
\includegraphics[scale=0.35, trim={2.5cm 0 0 0},clip]{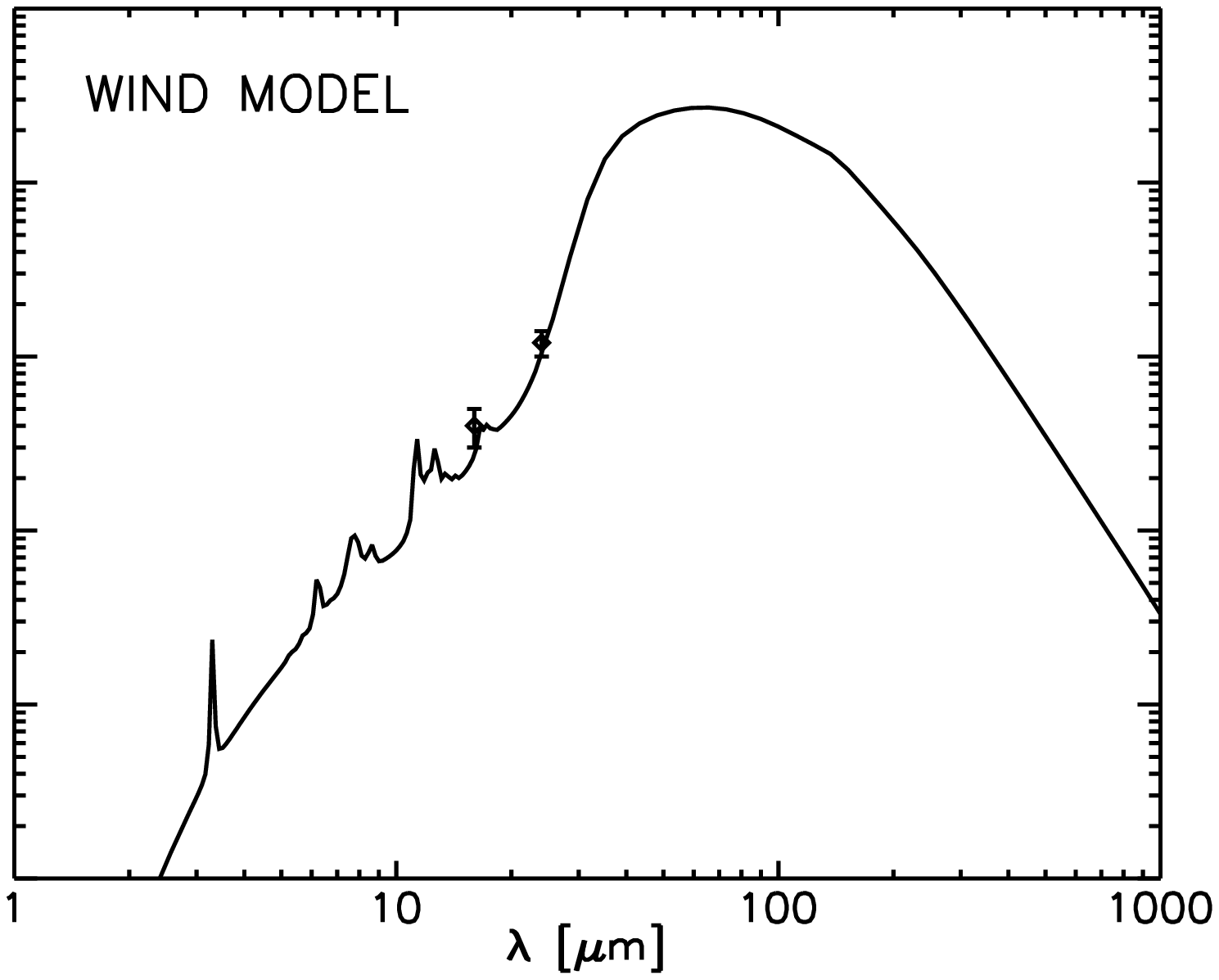}
\caption{Dust emission SED of the ring for the elliptical shell model (left panel) and the elliptical cavity model (central panel) and for the elliptical wind model (right panel) for two values of the gas density ($n_H=100-1000~\rm{cm^{-3}}$) and b/a=2.}
\label{fig_magtar_sed}
\end{figure*}

We note that also for the wind model we do not recover the ring morphology in the case the stars are located outside the cavity region. The dust emission SED of the ring predicted in the case the stars are located inside the cavity region is shown in Fig.\ref{fig_magtar_sed} (right panel). The model is able to fit the total fluxes at 16 and 24$\mu$m (see also integrated fluxes in Table \ref{table_tot_mid_fluxes}). This result confirms that a density of $n_H\sim1000 ~\rm{cm^{-3}}$ at the position of the border of the cavity region is necessary to fit the data, independently of the assumed dust distribution.

\subsection{Comparison of the average surface brightness profiles for the high density models}

\begin{figure}[t]
\includegraphics[scale=0.5,trim = 0 1cm 0 0, clip]{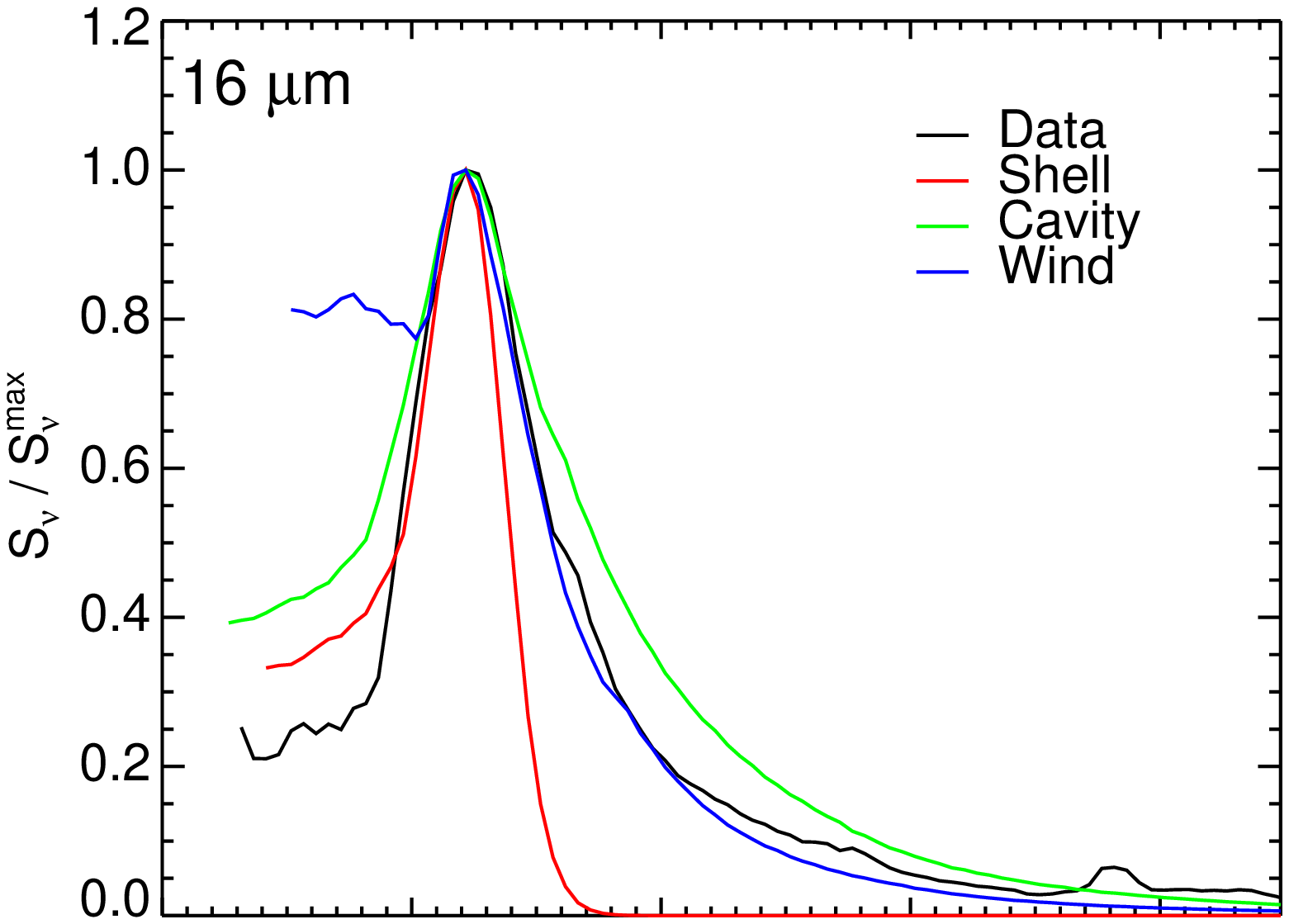}
\includegraphics[scale=0.5,clip]{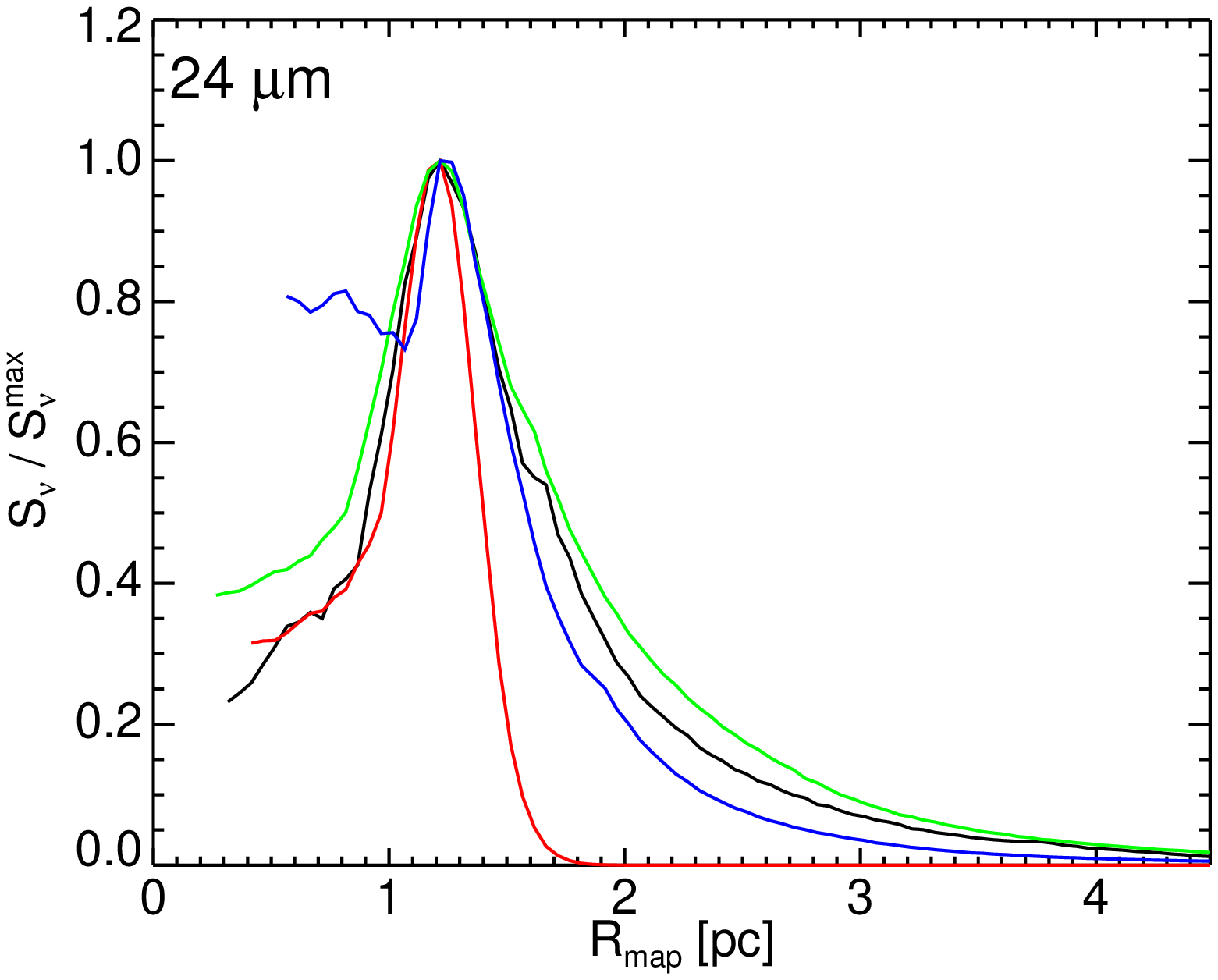}
\caption{Normalized surface brightness profiles of the data and the $n_{\rm H} = 1000$\,cm$^{-3}$ models with stars inside the cavity. The profiles have been derived by averaging over elliptical rings with same axis ratio and orientation as the infrared ring on the data maps.    }
\label{fig:norm_prof}
\end{figure}

In order to gain more quantitative insight on the similarities in morphology between the ring-like emission on the data and on the model maps, we compared average surface brightness profiles derived from the maps. We limited this comparison with the data to the models with the highest density (that is, $n_H=1000~\rm{cm^{-3}}$), which are the only ones able to reproduce the total ring emission flux (see above). We derived the average surface brightness profiles in the following way. Firstly, we masked the emission from the brightest stars on the data maps and the corresponding regions on the model maps. Then, from the background-subtracted maps, we derived average profiles by averaging the values of the non-masked pixels within elliptical rings with same centre, axis-ratio and orientation as the infrared ring on the data. Finally, we normalized all profiles to their maximum value. The surface brightness profiles so derived at 16 and 24\,$\mu$m are shown in Figure \ref{fig:norm_prof} for the data and the high density RT models for all the assumed dust density distributions. The radial distance R$_{\rm map}$ on the x-axis corresponds to the length of the semi-major axis of the elliptical rings used to derive the profiles. \\
From both the 16 and 24\,$\mu$m profiles several interesting features are evident. For radii larger than $\sim$1.2\,pc, the profile for the shell dust distribution ends too sharply and it is thus unable to reproduce the tail of the emission observed on the data, which is much more extended. On the other hand, the tails of the profiles for the cavity and wind dust distributions decrease in a smoother way that is much closer to that observed on the data (see also average discrepancies in table \ref{table_discrepancies_profiles}). This finding implies that a thin dust shell is not able to reproduce the ring emission profile. Thus, the presence of a large amount of dust beyond the outer edge of the cavity is required. Furthermore, the comparison for the inner part of the emission profiles (R$\leq$1.2\,pc) shows that the wind dust distribution gives rise to excess emission inside the ring, which is incompatible with the data. A sharp rise of the dust distribution at small radii provides much better agreement with the data (see table \ref{table_discrepancies_profiles}). The preferred dust distribution is of an almost dust-free cavity with a sharp transition to a dust rich environment. Beyond the cavity the dust and/or interstellar medium densities decrease as a function of distance slower than the $R^{-2}$ wind-model.

\begin{table}[t]
\caption{Average discrepancies for the model profiles in Fig.\ref{fig:norm_prof}$^{*}$}              
\label{table_discrepancies_profiles}      
\centering                                      
\begin{tabular}{l |c c| c c}          
\hline\hline                        
 Model & \multicolumn{2}{c}{$\Delta$(16\,$\mu$m)} & \multicolumn{2}{c}{$\Delta$(24\,$\mu$m)} \\ 
  & $R_{\rm map}<$1.2 & $R_{\rm map}>$1.2 & $R_{\rm map}<$1.2 & $R_{\rm map}>$1.2 \\
\hline                                   
SHELL & 4.2 & 7.1 & 1.5 & 8.1 \\
CAVITY&  7.8 & 2.6 & 4.7 & 1.5 \\
WIND   &  18 & 1.2 & 14 & 2.2 \\
\hline                                             
\end{tabular}
\begin{list}{}{}
\item[$^{*}$] {\it The tabulated discrepancies are calculated with respect to the data surface brightness profiles and are in units of the data background noise $\sigma$. The averages are shown separately for $R_{\rm map}<$1.2\,pc and $R_{\rm map}>$1.2\, pc.}
\end{list}
\end{table}

\section{Discussion}
\label{discussion}

We have performed several dust RT calculations assuming elliptical
dust shell/cavity geometries as well as a disrupted wind profile, and
by positioning the two supergiants stars inside or outside the dust
cavity. We have found that the dust ring morphology, similar to that
found on the Spitzer data of SGR1900+14, is recovered only in the
cases where the stars are inside the cavity. Furthermore, we
approximately reproduce the total integrated fluxes at 16 and
24\,$\mu$m only by assuming a gas density of
$n_H\sim1000~\rm{cm^{-3}}$ for all the dust geometries we assumed.
The corresponding mass of the dust responsible for the ring emission
is $M_{\rm dust}\sim$2\,M$_\odot$. \\

Given these results, the first question to ask is whether or not the
models that reproduce the observed dust emission morphology and total
flux are realistic or not. In particular the gas density, implied by
our modelling to explain the {\em Spitzer} infrared luminosity by dust
illumination, appears to be very high compared to that of the diffuse
galactic ISM.  Before discussing the possible nature of this high
density, we first clarify what assumptions/parameters in our modelling
might have caused an artificial high gas density, not representative
of the real ISM density around the magnetar. Firstly, we point out
that the gas density is not measured directly from the gas emission
but inferred from the dust density divided by a dust-to gas ratio of
0.00619 (which is characteristic of the assumed Milky Way dust model,
see section \ref{simulations}). However, this dust-to-gas ratio,
representative of the kpc scale ISM of the nearby Milky Way regions,
presents significant local variations in the ISM \citep[see
  e.g.][]{Reach15}. Furthermore, since the assumed size distribution
of the grains is also representative of the local Milky Way, this also
has an effect on the derived dust density. In fact, the MIR emission
in our modelling is mainly produced by small grains (sizes
$a\sim10^{-3}-10^{-2}~\mu$m) which are stochastically heated. If the
grain size distribution is more skewed towards smaller grain sizes,
compared to the one we are assuming, this would require significantly
less dust mass to reproduce the observed MIR emission. The grain size
distribution is known to be affected by both dust destruction and
formation processes, but it is not possible to constrain it further
with our observations. On the other hand, we also note that if the
cavity has been created by dust destruction, the grain size
distribution there should instead favour the presence of large dust
grains rather than small ones \citep{Waxman00, Perna02}.  In
  fact, a number of studies \citep{Fruchter01, Perna02, Perna03} have
  shown that the X-ray flux is more effective at destroying small
  grains than larger ones. The precise evolution of the dust grain
  distribution is dependent on both the spectral shape and overall
  intensity of the illuminating source, on the composition of the
  grains, as well as on the relative importance of the processes of
  X-ray Heating, Coulomb Explosion and Ion Field Emission, the last
  two of which being particularly uncertain \citep{Fruchter01,
    Perna02}. However, even within these uncertainties, all the models
  generally predict that smaller grains will be destroyed to larger
  distances than larger ones\footnote{If Coulomb Explosion dominates,
    then the distance to which grains of size $a$ are destroyed by a
    source with energy spectral index $\alpha$ scales as
    $a^{-0.5-\alpha/3}$ if grain charging is limited by internal
    energy losses, or as $a^{-1-\alpha/2}$ if limited by the
    electrostatic potential.}. Therefore,  there would be a region in which only selective destruction took place, leaving behind a dust distribution skewed towards large grains. At the inner edge the distribution would be skewed towards big grains, progressively changing into the undisturbed (pre-burst) distribution at larger distances.     
  An attempt at modeling these
  effects would be worthwhile if the quality of the data were to allow
  a comparison with observations, but this is not possible with the
  current data.

Finally, in our modelling we only assumed the two supergiant stars,
the most luminous stars in the field, to be heating the dust
shell/cavity. However, other sources of radiation might well play a
role (e.g. other fainter stars within the cavity) and, in this case,
the needed gas density to match the observed fluxes would be lower. On
the other hand, note that the constant $\sim10^{34}$\,erg/s X-ray
luminosity emitted by the magnetar is too low to power the dust
emission. In fact, the wavelength--integrated dust emission luminosity
for the models that fit the MIR fluxes is in the range 3.7-4 $\times
10^{35}$\,erg/s. An additional mechanism to heat the dust is also
collisional heating in hot plasma, where the dust is heated by the
collisions with high energy electrons. This is expected if the dust is
embedded in shocked gas with temperatures of order of 10$^6$
K. However, in this case we might expect to see an X-ray diffuse
emission from the hot gas around the magnetar as well, which is not
observed.\\

Vrba et al.(2000) argued that SGR 1900+14 is associated with a cluster
of young stars (much fainter in apparent magnitude than the two M
supergiants we considered in our modelling) which are probably
embedded in a dense medium. This interpretation is qualitatively
consistent with our results.  As proposed by \citep{Wachter08}, the
1998 Giant Flare could have produced the cavity by destroying the dust
within it. Assuming a constant dust density within the cavity region,
corresponding to n$_H$=1000\,cm$^{-3}$, we estimated a total dust mass
of order of 3\,M$_\odot$ that was plausibly present before being
destroyed by the flare. An energy of about $E\sim 6\times10^{45}$erg
would suffice to destroy this amount of dust, consistent with the
estimates by \citep{Wachter08} based on Eq. 25 in
\citet{Fruchter01}.  The size of the region with destroyed dust
would be larger for smaller grains, as discussed above.
In this scenario, the high density we
derived would be similar to the high density ISM around the
magnetar. Furthermore, we note that high density of the ISM
(n$_H$=10$^5$--10$^7$\, cm$^{-3}$) has been found in the environment
surrounding GRBs \citep{Lazzati02}, which should be similar to that
where magnetars are located.

The wind model we considered was meant to be similar to the scenario where the dust distribution outside the cavity was mainly determined by the wind of the magnetar progenitor while internally disrupted by the Giant Flare. However, gas densities at 1pc distance in a typical stellar wind are expected to be several orders of magnitudes lower than those we found ($n_H\sim1$\,cm$^{-3}$,  estimated from the mass-loss rates $\dot{M}$ of \citet{Kudritzki02} assuming $n_H \propto \dot{M}/v/r^2$ with $v$ = 1000 km/s). Thus, this last scenario is unlikely if the ring density is indeed so high.  \\

Another possibility is that the dust emission ring is the infrared emission from the supernova remnant (SNR) of the magnetar progenitor. The dust mass associated with the shell model with $n_H$=1000\,cm$^{-3}$ is $M_{dust}$=1.9\,M$_\odot$, which is a factor 3--4 higher than the measurement of dust mass around SN1987 by \citet[0.4--0.7 $M_\odot$]{Matsuura11}. However, given the large uncertainties in the inferred dust masses, and that our value for the dust density might have been overestimated because of the reasons given above, it may well be that the amount of dust needed for the shell model is compatible with that of SNRs. The SNR scenario was also considered by \citep{Wachter08} but discarded because of the lack of observed radio and X-ray emission from the ring. However, if we consider i) the IR/X luminosity ratio of $\sim10^{-1}-10^{2}$ measured by \citet{Koo16} for many SNRs, ii) the total IR luminosity of the ring in our models ($\sim4\times10^{35}$\,erg/s), and the X-ray detection limit for the ring \citep[$\sim2\times10^{33}$erg/s in the 2--10keV band][]{Wachter08}, this structure is still compatible with being a SNR with high IR/X ratio.\\
On the other hand, we can also compare the 24$\mu$m luminosity with the expected X-ray luminosity, according to Figure 12 of \citet{Seok13}, which studied a sample of SNR in the Large Magellanic Cloud. If the magnetar was located at the distance of the LMC (50\,kpc), we would have $\nu F_\nu(24\mu m)= 1.5\times10^{-10}$\,erg/s/cm$^2$, and the relative expected X-ray flux would be $2\times10^{-10}$\,erg/s/cm$^2$. This would translate in an intrinsic X-ray luminosity of $\sim6\times10^{37}$\,erg/s, which should have been clearly detected in the case of the SGR 1900+14. Hence, given the information we have at hand we cannot discard the SNR scenario on the base of the observed IR/X-ray luminosity ratio, although it would be a rather peculiar remnant compared with what we see around other Galactic pulsars or magnetars \citep{Green84, Martin14}. 

However, if we also consider the shape of the normalized average surface brightness profiles, shown in Fig.\ref{fig:norm_prof}, this provides a strong evidence that 1) there is very little amount of dust inside the cavity and 2) the emitting dust is much more extended than a simple thin shell. These findings are compatible with the scenario where the cavity has been produced by the Giant Flare within a high density medium. However, the SNR scenario would still be acceptable in the case the transition in density between the shell and the surrounding ISM is smoother than what we assumed in our modelling. 

Regardless of the origin or the exact distribution of the illuminated dust, or the exact nature of the dust free cavity, our models show that we are able to observe this illuminated dust structure only because of two favourable characteristics: 1) the high dust density in the local region, and 2) the illuminating stars coincidentally lay inside the shell. Similar dust structures might potentially be present around many other magnetars or pulsars but they would be invisible to us because of the lack of either one of the two above local properties of this particular object.

\acknowledgments

GN acknowledges support by the EU COST Action MP1304, for the Short Term Scientific Mission where this project was completed, and by the Leverhulme Trust research project grant RPG-2013-41. N.R. acknowledges funding in the framework of the NWO Vidi award A.2320.0076, and via the European COST Action
MP1304 (NewCOMPSTAR). N.R. and D.F.T. are supported by grants AYA2015-71042-P and SGR2014-1073.  RP acknowledges support from the NSF under grant AST-1616157. JMG is supported by grant AYA2014-57369-C3-1-P.

\end{document}